\documentstyle[prd,aps,epsf,12pt]{revtex}

\def\leftvec{\raise1.5ex\hbox{$\leftarrow$}\kern-.85em}
\def\half{{\scriptstyle \raise.15ex\hbox{${1\over2}$}}}
\def\threehalves{{\scriptstyle \raise.15ex\hbox{${3\over2}$}}}
\def\third{{\scriptstyle \raise.15ex\hbox{${1\over3}$}}}
\def\twothirds{{\scriptstyle \raise.15ex\hbox{${2\over3}$}}}
\def\fourthirds{{\scriptstyle \raise.15ex\hbox{${4\over3}$}}}
\def\fourth{{\scriptstyle \raise.15ex\hbox{${1\over4}$}}}
\def\gtwid{\raise.3ex\hbox{$>$\kern-.75em\lower1ex\hbox{$\sim$}}}
\def\ltwid{\raise.3ex\hbox{$<$\kern-.75em\lower1ex\hbox{$\sim$}}}

\def\ie{{\it i.e.},\ }
\def\eg{{\it e.g.},\ }

\def\et{{\it et al.}}

\def\cI{{\cal I}}

\def\cL{{\cal L}}
\def\cM{{\cal M}}

\def\cO{{\cal O}}

\def\cV{{\cal V}}

\def\eq#1{eq.~(\ref{eq:#1})}
\def\eqs#1#2{eqs.~(\ref{eq:#1}) and (\ref{eq:#2})}
\def\eqsthree#1#2#3{eqs.~(\ref{eq:#1}), (\ref{eq:#2}) and (\ref{eq:#3})}
\def\eqsfour#1#2#3#4{eqs.~(\ref{eq:#1}), (\ref{eq:#2}), (\ref{eq:#3}) and (\ref{eq:#4})}

\def\prd#1{Phys.\ Rev.\ {\bf D#1}}
\def\plb#1{Phys.\ Lett.\ {\bf #1B}}
\def\npb#1{Nucl.\ Phys.\ {\bf B#1}}

\def\seillac{Nucl.\ Phys.\ {\bf B} (Proc.\ Suppl.) {\bf 4} (1988)}

\def\capri#1{Nucl.\ Phys.\ {\bf B} (Proc.\ Suppl.) {\bf 17}, #1 (1990)}

\def\berlin{presented at the International Symposium,
{\it Lattice 2001}, Berlin, August 19--24, 2001, to be published
in Nucl.\ Phys.\ {\bf B} (Proc.\ Suppl.)}

\def\MeV{{\rm Me\!V}}
\def\GeV{{\rm Ge\!V}}

\begin{document}

\draft
\title{
Chiral Logs in the Presence of Staggered \\
Flavor Symmetry Breaking
\medskip
}
\author{
C.~Bernard
}
\address{
(MILC Collaboration)\\
\smallskip
Washington University, St.~Louis, Missouri 63130, USA\\
\smallskip
}
\date{\today}

\maketitle


\begin{abstract}\noindent
Chiral logarithms in $m_\pi^2$ are calculated at one loop, taking
into account the leading contributions to flavor symmetry breaking due
to staggered fermions.  I treat both the full QCD case (2+1 light
dynamical flavors) and the quenched case; finite volume corrections
are included.  My starting point is the effective chiral
Lagrangian introduced by Lee and Sharpe.
It is necessary to understand the one-loop diagrams in the ``quark flow''
picture in order to adjust the calculation to correspond to the desired
number of dynamical quarks.  
\end{abstract}

\section{INTRODUCTION}
\label{sec:intro}
Staggered (Kogut-Susskind, KS) fermions provide a competitive
method for simulating full QCD, including the effects
of virtual $u$, $d$ and $s$ pairs. Indeed, recent
improvements in the action \cite{IMP_KS} have made it possible
to compute many physical quantities with rather small scaling
violations 
in such ``2+1 flavor" simulations \cite{IMP_SCALING,MILC_SPECTRUM}.

The violation of KS-flavor symmetry, while
reduced by the improved (``Asqtad'') action, remains however quite large at
the lattice spacing ($a\approx 0.13$ fm) where most of the 
simulations in Refs.~\cite{IMP_SCALING,MILC_SPECTRUM} have been performed.
The maximum splitting in mass squared among the
various flavor pions, $\Delta_{\rm max}$,
is  $\approx (400\; \MeV)^2$ at this lattice spacing. 
Because one can choose the staggered flavor on the valence quark
lines, the flavor violations often enter through 
loop effects alone, and as such they have  a typical size of only
$\sim$8\% in most quantities.  Here, I  estimate  the size from
a ``typical'' chiral logarithm, including flavor violation:
\begin{equation}
\label{eq:DELTAMAX}
{\Delta_{\rm max}\over 16 \pi^2 f_\pi^2} \ln (\Delta_{\rm max}/\Lambda^2)\ ,
\end{equation}
where $\Lambda$ is the chiral scale
(taken, for instance equal to $m_\rho \approx 770\; \MeV$),
and $f_\pi\cong 131\; \MeV$. Since the
leading flavor violating terms in the improved KS action are $\cO(a^2)$,
these discretization effects can be reduced still more by going closer
to the continuum limit; MILC simulations are currently in progress
at $a\approx 0.09$ fm.

On the other hand, if one focuses directly
on chiral logs and works at $a\approx 0.13$ fm, the effect of 
KS-flavor violation should be large.  Indeed,
in Ref.~\cite{MILC_SPECTRUM}, we were unable to fit
$m_\pi^2/m_\ell$  to the standard continuum chiral-log 
form \cite{GASSER_LEUTWYLER} for 2+1 flavor QCD.
(Here and below $\ell$ stands for a generic light ($u$ or $d$) quark; I neglect
isospin violations throughout.)
To test the hypothesis that KS-flavor violations are responsible
for this behavior of $m_\pi^2/m_\ell$, one needs to compute
the chiral logs in the presence of KS symmetry breaking.
That computation is the subject of this paper \cite{ORGINOS}.

The effective chiral Lagrangian that describes
KS fermions through $\cO(m,a^2)$ ($m$ is a generic quark mass) has been 
constructed by Lee and Sharpe \cite{LEE_SHARPE}.  Their approach is to make
a joint expansion in $m$ and $a^2$, which are considered to be comparably small.
That is in fact the case for a simulation like \cite{MILC_SPECTRUM}, where
the splittings in mass squared of various KS-flavor pions are comparable to
the squared masses themselves. 
A one-loop calculations using the Lagrangian of Ref.~\cite{LEE_SHARPE}
would give the chiral logs (non-analytic terms) of $\cO(m^2,ma^2,a^4)$.   

The Lee-Sharpe Lagrangian corresponds to
a single lattice KS field, which becomes 4 degenerate flavors
in the continuum limit.
The main subtleties in the current work arise because I am interested in the 
more phenomenologically relevant 2+1 flavor theory being simulated,
with $m_s \not= m_\ell$ in general,
not the 4-flavor (continuum degenerate) KS theory. 
 I follow a three-step procedure:

\begin{itemize}

\item[]{1.} Generalize (almost trivially) the Lee-Sharpe Lagrangian to
the 4+4 case, where one has two lattice KS fields with different
masses ($m_\ell$, $m_s$).  This is an $SU(8)_L \times SU(8)_R$ theory,
with mass terms and $\cO(a^2)$ KS-flavor violating 
terms breaking the symmetry.

\item[]{2.} Compute $m_\pi^2$ at one loop in the 4+4 case.  
I treat only the case where the pion is the $U(1)_A$  Goldstone
particle. (This is the
situation most immediately relevant to the fits attempted in
\cite{MILC_SPECTRUM}.) The symmetry then implies that
$m_\pi^2$ vanishes when $m_\ell\to 0$, allowing one to
simplify  the calculation.

\item[]{3.}  Adjust the 4+4 answer to correspond to the 2+1
case of interest. This requires identifying the contributions
that correspond to $\ell$ or $s$ virtual quark loops
and multiplying them by $1/2$ or $1/4$ respectively. 

\end{itemize}

The subtleties are almost entirely in step 3. 
To identify those terms that should be reduced by
a factor of 2 or 4, one needs to follow 
the ``quark flow'' approach, which was introduced
by Sharpe \cite{SHARPE_QCHPT} to compute quenched chiral logarithms.
Unfortunately, in the presence of KS-flavor
symmetry violation, there does not appear to be an alternative, Lagrangian
approach \cite{CBMG_QCHPT,CBMG_PQCHPT} that would
automatically cancel the effects of the unwanted KS flavors.  
This corresponds to the fact that, 
in numerical simulations, the reduction from 4+4
to 2+1 flavors is accomplished by taking the square root
and fourth root of the $\ell$ and $s$ determinants, respectively.
The procedure reduces each of the four KS flavors in a virtual quark
loop equally, by a factor of 2 (4); it does not than cancel specific
flavors.  There is thus no equivalent (ultra)local lattice Lagrangian, and 
it is not at all clear how or whether one can represent
the low energy chiral properties of this theory with an
effective chiral Lagrangian \cite{HASENFRATZ}.  

Of course, in
the continuum limit the KS-flavor symmetry is restored, and one
may cancel any 2 of the 4 light quark flavors (and any 3 of
the 4 strange quark flavors) with 2 (3) bosonic pseudo-quarks. The effective
chiral theory is then obvious:
an ``$8|5$'' partially quenched chiral perturbation theory.
This is discussed
in Ref.~\cite{CBMG_PQCHPT}.  Such a theory is trivially equivalent
to standard 3 flavor chiral perturbation theory in the physical
sector where the sea quarks are also the ones on the external lines.

The remainder of this paper is organized as follows:
Section \ref{sec:LAGRANGIAN} presents the Lee-Sharpe Lagrangian and
generalizes it to the 4+4 case.  The calculation of $m_\pi^2$
to one loop is then described in Sec.~\ref{sec:CALCULATION}. 
In Sec.~\ref{sec:FLOW}, the quark flow
picture is applied to compute $m_\pi^2$ in the 2+1 case. The analytic
terms at $\cO(m^2,ma^2,a^4)$, necessary to perform consistent fits that
include the chiral logs, are also touched upon.  The quenched
case follows in Sec.~\ref{sec:QUENCH}. 
Finite volume corrections are then discussed
briefly in Sec.~\ref{sec:FINITE_VOLUME}.
Finally, I make some remarks about the
extension of the current approach to other physical quantities
in Sec.~\ref{sec:CONCLUSION}.  That section also contains some
comments about the fits of MILC data to the chiral forms derived
here; however, a detailed discussion of such fits and their consequences
will be presented elsewhere \cite{MILC_FITS}.

\section{Lee-Sharpe Lagrangian for 4+4 Flavors}
\label{sec:LAGRANGIAN}

Lee and Sharpe \cite{LEE_SHARPE} first construct the
continuum effective Lagrangian for  KS lattice fermions, including
all terms at $\cO(a^2)$.  The KS-flavor symmetry is broken by
6-quark operators.  They then write down the corresponding
effective chiral theory to leading order in $a^2$ and $m$,
the quark mass.

For a single KS field (4 continuum flavors) the Lee-Sharpe chiral
Lagrangian describing the pseudo-Goldstone bosons has a (nonlinear)
$SU(4)_L\times SU(4)_R$ symmetry, broken by the mass term and by
the KS-flavor violating operators. The  
4$\times$4 matrix $\Sigma$ is defined by
\begin{equation}
\Sigma(x) \equiv \exp(i\phi/f)\  ,  \qquad \phi \equiv \sum_{a=1 }^{16}\phi_a T_a \,
\label{eq:SIGMAandPHI}
\end{equation}
where $\phi_a$ are real, $f$ is the tree-level pion decay constant
(I use the normalization $f_\pi\cong131\ \MeV$), and the 
Hermitian generators $T_a$ 
are chosen as follows,
\begin{equation}
T_a = \{\xi_5,i\xi_{\mu5},i\xi_{\mu\nu},\xi_\mu,I\} \ .
\label{eq:T}
\end{equation}
Here $\xi_\mu$ are the flavor gamma matrices, 
$\xi_{\mu5}\equiv \xi_\mu\xi_5$, 
$\xi_{\mu\nu}\equiv \xi_\mu\xi_\nu$ 
(with $\mu<\nu$ in eq.~(\ref{eq:T})), 
and $I$ is the 4$\times$4 identity matrix.  
$\Sigma$ transforms by $\Sigma\to L\Sigma R^\dagger$ under 
$SU(4)_L\times SU(4)_R$.

Note that I include the singlet meson $\phi_I$,
corresponding to generator $I$, in eqs.~(\ref{eq:SIGMAandPHI}) and (\ref{eq:T}).  
The mass of the $\eta'$-like meson 
(often called $\Phi_0$, up to overall normalization)
gets a large 
contribution ($\equiv m_0$)
from the anomaly in the Lee-Sharpe case and is not included
in their formalism.
But I want to generalize to two KS fields
(8 continuum flavors).  In that case, only the $SU(8)$ flavor singlet
is heavy and can be integrated out. $SU(4)$ singlets,
\eg a singlet ``kaon'' made of one quark from each KS multiplet,
can be light.  I therefore leave the singlet as well
as the $m_0$ term explicit in intermediate steps, but will ultimately take
$m_0\to\infty$.\footnote{Sharpe and Shoresh \cite{SHARPE_SHORESH}
show that integrating out the $\Phi_0$ is mathematically equivalent to
keeping it in the calculations and 
taking the mass parameter $m_0\to\infty$
at the end.  If that limit is to be taken, it is unnecessary to include
$\Phi_0$ dependence other than the $m_0$ term in the action.
However, in the quenched case (Sec.~\ref{sec:QUENCH}), other 
$\Phi_0$ dependence
will need to be considered.}

The (Euclidean) Lee-Sharpe Lagrangian is then
\begin{equation}
\cL_{(4)} = {f^2\over 8} {\rm tr}(\partial_\mu\Sigma 
\partial_\mu \Sigma^\dagger)
-{1\over4}\mu m f^2 {\rm tr}(\Sigma + \Sigma^\dagger) + {2m_0^2\over3}(\phi_I)^2 + a^2\cV \ , 
\label{eq:LS_LAG} 
\end{equation}
where $\mu$ is a constant with units of mass, 
the coefficient of $\phi_I$ is conventional,\footnote{\ $m_0^2/6$ by 
definition multiplies $(\phi_{11}+\phi_{22}
+\phi_{33}+\cdots)^2$. Here, the
normalized $\phi_I\equiv 
\half\sum_{i=1}^4 \phi_{ii}$.  This definition of $m_0^2$ is independent
of the number of flavors and corresponds to that
in Refs.~\cite{CBMG_QCHPT,CBMG_PQCHPT} (after renaming $\mu^2\to m_0^2$)
and that in Ref.~\cite{SHARPE_QCHPT} (after choosing $N=3$ there).
With the current definition of $f$, the parameter $\delta$ introduced in 
\cite{SHARPE_QCHPT} is equal to $m_0^2/(24\pi^2 f^2)$.}
and $\cV$ is the lowest order KS-flavor breaking potential:
\begin{eqnarray}
-\cV &= &C_1\,  {\rm tr}(\xi_5\Sigma \xi_5 \Sigma^\dagger)\nonumber \\
&&+ C_2\, \half\big[{\rm tr}(\Sigma^2) - {\rm tr}(\xi_5\Sigma \xi_5 \Sigma) + h.c.\big]\nonumber \\
&&+ C_3\, \half\sum_\nu\big[{\rm tr}(\xi_\nu\Sigma \xi_\nu \Sigma) + h.c.\big]\nonumber \\
&&+C_4\, \half\sum_\nu\big[{\rm tr}(\xi_{\nu5}\Sigma \xi_{5\nu} \Sigma) + h.c.\big]\nonumber \\
&&+ C_5\, \half\sum_\nu\big[{\rm tr}(\xi_\nu\Sigma \xi_\nu \Sigma^\dagger) -
tr(\xi_{\nu5}\Sigma \xi_{5\nu} \Sigma^\dagger) \big]\nonumber \\
&&+ C_6\, \sum_{\mu<\nu}\big[{\rm tr}(\xi_{\mu\nu}\Sigma \xi_{\nu\mu} \Sigma^\dagger) \ .
\label{eq:V}
\end{eqnarray}
Effects of order $a^2m$ or $a^4$ are neglected in eq.~(\ref{eq:LS_LAG}).

The potential $\cV$ has an ``accidental'' $SO(4)$ flavor symmetry (evidenced
by the sums over $\nu$ and $\mu$), which is larger than the lattice
symmetry group.  The pions fall into 5 $SO(4)$ representations with flavors
$\xi_5$, $\xi_{\mu5}$, $\xi_{\mu\nu}$, $\xi_\mu$, $I$.  Expanding
eq.~(\ref{eq:LS_LAG}) to quadratic order in $\phi$, the pion masses
are found to be:
\begin{equation}
m^2_{\pi_B} = 2\mu m + {4m_0^2\over3}\delta_{B,I}+ 
               a^2\Delta(\xi_B)  \ ,
\label{eq:MASS4}
\end{equation}
where, in a convenient abuse of notation that will be used
from now on,  the index
$B$ takes the 16 values $\{5,\mu5,\mu\nu (\mu<\nu),\mu, I\}$
and $\xi_I\equiv I$.  
The $m_0^2$ contribution to the mass of $\phi_I$ is shown
explicitly, and  
$\Delta(\xi_B)$ is given by
\begin{eqnarray}
\Delta(\xi_5) &=& 0 \nonumber \\
\Delta(\xi_{\mu5}) &=& {16\over f^2}(C_1 + C_2 + 3C_3 + C_4 - C_5 + 3C_6) 
\nonumber \\
\Delta(\xi_{\mu\nu}) &=& {16\over f^2}(2C_3 + 2C_4 + 4C_6) 
\nonumber \\
\Delta(\xi_{\mu}) &=& {16\over f^2}(C_1 + C_2 + C_3 + 3C_4 + C_5 + 3C_6) 
\nonumber \\
\Delta(I) &=& {16\over f^2}(4C_3 + 4C_4 ) 
\label{eq:DELTAS}
\end{eqnarray}
$\Delta(\xi_5)$ vanishes because of the staggered-flavor nonsinglet 
$U_A(1)$, which is 
represented in the chiral theory by
\begin{equation}
\Sigma \to e^{i\theta\xi_5} \Sigma e^{i\theta\xi_5},
\label{eq:UA1} 
\end{equation}
with $\theta$ the $U_A(1)$ angle.  Since this
symmetry is unbroken by the lattice regulator,
$\pi_5$ is a true Goldstone boson in the chiral limit.

From simulations such as those in Ref.~\cite{MILC_SPECTRUM}, one learns that
the 
$C_4$ term in eq.~(\ref{eq:V})
is the largest contributor to the $\cO(a^2)$ flavor 
violation \cite{LEE_SHARPE}. 
This leads to approximately equal splitting between the pions, in the 
order
$\pi_5$, $\pi_{\mu5}$, $\pi_{\mu\nu}$, $\pi_\mu$, $\pi_I$.\footnote{
Omitting the disconnected terms in the $\pi_I$ propagator
(as is often done in simulations -- see, \eg \cite{MILC_SPECTRUM}),
eliminates the $m_0$ contribution to the singlet mass and extends
the approximately equal splitting rule to KS flavor $I$.}     
The other operators  are not entirely absent, however; 
their contributions to the splittings 
are of order $10\%$ of that of $C_4$ \cite{MILC_FITS}.
The reason for the smallness of $C_i$, $i\not=4$, is not known.\footnote{
More precisely, the approximately equal splittings imply only
that $C_6$, $C_1+C_2$, and $2C_3-C_5$ are small.  I thank M.\ Golterman 
for pointing this out.}

We now need to generalize 
to the case of two KS fields with different masses $m_\ell$ and $m_s$,
\ie 4+4 continuum flavors.  The field $\Sigma$ in eq.~(\ref{eq:SIGMAandPHI})
becomes an $8\times8$ matrix, given by
\begin{equation}
\Sigma(x) \equiv \exp(i\Phi/f)\ ,  
\qquad \Phi \equiv \left(\matrix{\pi&K\cr
K^\dagger&S\cr}\right) \ ,
\label{eq:SIGMAandPHI88}
\end{equation}
where the $4\times4$ fields $\pi$, $K$, and $S$ 
describe ``pions,''  ``kaons,'' and  ``$s\bar s $'' mesons, respectively.
As in eqns.~(\ref{eq:SIGMAandPHI},\ref{eq:T}),
$\pi \equiv \sum_{a=1 }^{16}\pi_a T_a$, with $\pi_a$ real, 
and similarly for $K$ and $S$, except that the $K_a$ are complex.
  The $SU(8)$ singlet $\propto {\rm tr}\pi+{\rm tr}S$
will be eliminated below by taking the $m_0\to\infty$ limit.

Defining the $8\times8$ mass matrix $\cM$,
\begin{equation}
\cM \equiv \left(\matrix{ m_\ell I &  0 \cr
                         0    & m_s I\cr } \right)\ ,
\label{eq:M}
\end{equation}
the $SU(8)_L\times SU(8)_R$ Lagrangian that generalizes $\cL_{(4)}$ of
eq.~(\ref{eq:LS_LAG})
 is then
\begin{equation}
\cL_{(4+4)} = {f^2\over 8} {\rm tr}(\partial_\mu\Sigma 
\partial_\mu \Sigma^\dagger)
-{1\over4}\mu  f^2 {\rm tr}(\cM\Sigma + \cM\Sigma^\dagger)+ {2m_0^2\over 3}
(\pi_I + S_I)^2 + a^2\cV \ .
\label{eq:SU8_LAG} 
\end{equation}

Generalization of the KS-flavor breaking 
potential $\cV$ in eq.~(\ref{eq:V}) requires
a little thought. For a single KS field, the symmetry breaking
4-quark operators in the effective continuum theory
have the generic form 
\begin{equation}
\bar Q (\gamma_S \otimes \xi_F) Q\; 
\bar Q (\gamma_{S'} \otimes \xi_{F'}) Q\  ,
\label{eq:FOURQUARK} 
\end{equation}
where $\gamma_S$ is an arbitrary spin matrix,
$\xi_F$ is an arbitrary flavor matrix, and the
indices $S,S',F,F'$ are contracted in various ways determined
by the lattice symmetries. The effective chiral operators
may then be found from the 4-quark operators by treating
$\xi_F$ and $\xi_{F'}$ as spurion fields \cite{LEE_SHARPE}.

With two KS fields on the lattice, there
is an exact vector $SU(2)$ (broken only by $m_\ell\not=m_s$) that
mixes them.\footnote{There are also corresponding axial symmetries,
whose generators are the direct product of the $U_A(1)$ generator,
$\gamma_5 \otimes \xi_5$, with
the vector $SU(2)$ generators, but I will not need them here.}
The symmetry guarantees that the KS-flavor breaking 4-quark operators
now have either of the two forms:
\begin{equation}
\bar Q_i (\gamma_S \otimes \xi_F) Q_j\; 
\bar Q_j (\gamma_{S'} \otimes \xi_{F'}) Q_i\ , \qquad 
\bar Q_i (\gamma_S \otimes \xi_F) Q_i\; 
\bar Q_j (\gamma_{S'} \otimes \xi_{F'}) Q_j\ ,
\label{eq:FOURQUARK8} 
\end{equation}
where $i$ and $j$ ($=1,2$) are $SU(2)$ indices.  The operators
in eq.~(\ref{eq:FOURQUARK8}) are ``flavor mixed'' and ``flavor unmixed,''
respectively.  By Fierz transformation, the flavor mixed operators can
be brought to the flavor unmixed form, so we may assume
all 4-quark operators are of the latter type.\footnote{This may be at the expense
of mixing the color indices 
(suppressed in eqs.~(\ref{eq:FOURQUARK}) and (\ref{eq:FOURQUARK8})), but they
are irrelevant in the corresponding chiral effective theory because of
confinement. }  But if all 4-quark operators are flavor unmixed,
then all spurion fields are $SU(2)$ singlets. 
In other words, for $\cV$ in eq.~(\ref{eq:SU8_LAG}) we may take
simply eq.~(\ref{eq:V}) with the replacement
\begin{equation}
\label{eq:SU8_SPURION}
\xi_B \to 
\left(\matrix{ \xi_B &  0 \cr
                         0    & \xi_B \cr } \right)\ .
\end{equation}

The tree level masses of the pions, kaons, and $S$ ($s\bar s$) mesons
are then
\begin{eqnarray}
m^2_{\pi_B} &=& 2\mu m_\ell + a^2\Delta(\xi_B) \nonumber \\
m^2_{K_B} &=& \mu (m_\ell+m_s) + a^2\Delta(\xi_B) \nonumber \\
m^2_{S_B} &=& 2 \mu m_s + a^2\Delta(\xi_B) \ ,
\label{eq:MASS8}
\end{eqnarray}
where $\Delta(\xi_B)$ are given by eq.~(\ref{eq:DELTAS}),
and the $m_0$ terms  are not included because we will treat
$2m_0^2(\pi_I+S_I)^2/3$ in eq.~(\ref{eq:SU8_LAG}) as a vertex
(summed to all orders) below.

\section{One-loop pion mass for 4+4 Dynamical Flavors}
\label{sec:CALCULATION}

The calculation of $m_\pi^2$ to one loop in the 4+4 theory
(eq.~(\ref{eq:SU8_LAG})) is now straightforward.
I confine my attention to corrections
to the Goldstone pion mass $m_{\pi_5}$.  The cases of the non-Goldstone 
pions and the kaons, which will also be useful in confronting
simulation data, will be left to future publications.

\begin{figure}[thb]
\epsfxsize=0.99 \hsize
\epsffile{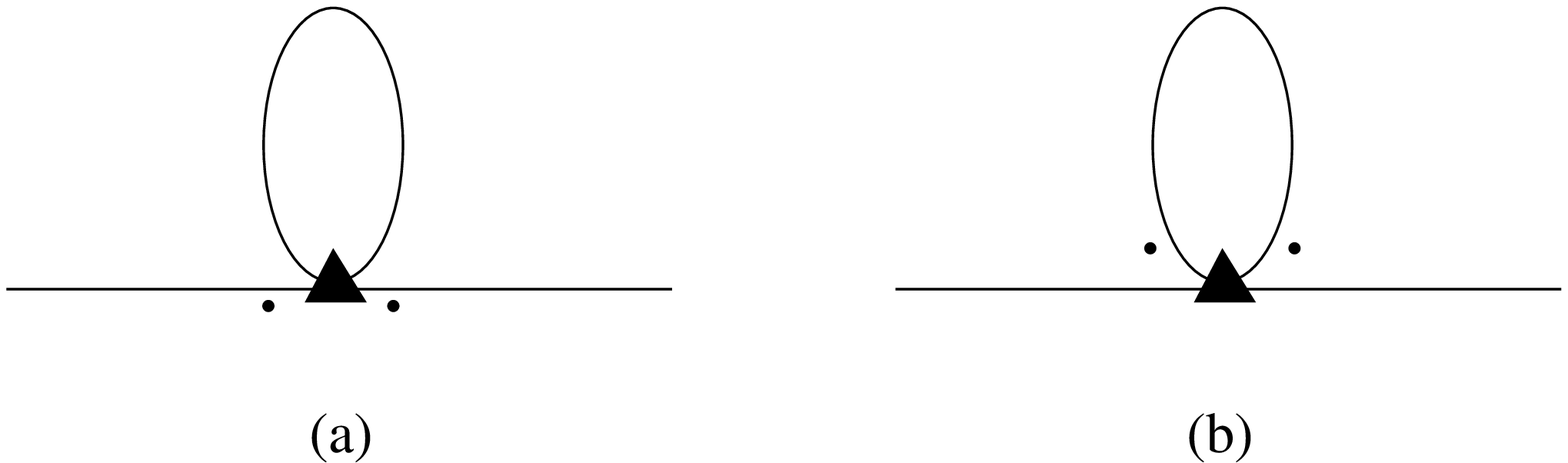}
\vspace{0.3truein}
\caption{Chiral perturbation theory graphs contributing to the pion propagator from
kinetic energy vertex (solid square). 
The external lines are Goldstone pions, \ie $\pi_5$.
The dots represent the derivatives in the vertex. 
In (a) they act on the external
lines; in (b), the internal. 
}
\label{fig:KE_vertex}
\end{figure}

\begin{figure}[thb]
\epsfxsize=0.99 \hsize
\epsffile{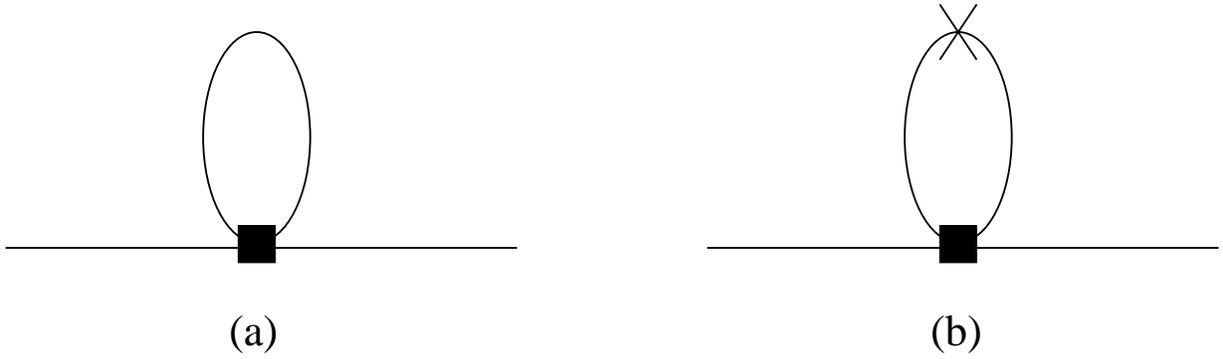}
\vspace{0.3truein}
\caption{Same as Fig.~\protect{\ref{fig:KE_vertex}}, but
from the mass vertex (solid triangle). 
The internal propagator in (a) is the connected propagator only
(no $m_0^2$ insertions),
even when it is neutral ($\pi_I$).
All disconnected contributions are in (b); \ie the cross represents
one {\it or more} insertions of the $m_0^2$ vertex.
}
\label{fig:MASS_vertex}
\end{figure}

\begin{figure}[thb]
\null
\vspace{0.5truein}
\epsfxsize=0.49 \hsize
\epsffile{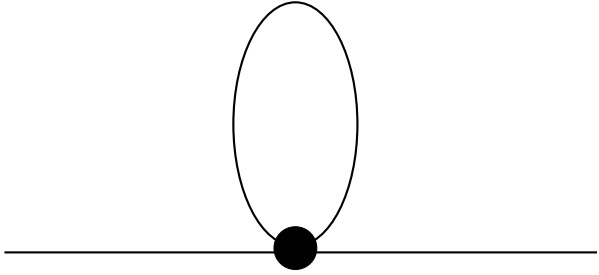}
\vspace{0.3truein}
\caption{Same as Fig.~\protect{\ref{fig:KE_vertex}}, but
from the flavor breaking  vertex $\cV$ (solid circle). 
}
\label{fig:V_vertex}
\end{figure}

The graphs are all tadpoles, with vertices coming from the 
kinetic energy term ${\rm tr}(\partial_\mu\Sigma
\partial_\mu \Sigma^\dagger)$,  the
mass term ${\rm tr}(\cM\Sigma + \cM\Sigma^\dagger)$,
and the
symmetry violating term $\cV$.  They are shown in 
Figs.~\ref{fig:KE_vertex}, \ref{fig:MASS_vertex}, 
and \ref{fig:V_vertex}, respectively.

I write the one-loop $\pi_5$ self energy as\footnote{The distinction
between the field matrix $\Sigma$ in the Lagrangian and the
self energy $\Sigma$ -- both conventional notations -- should be
clear from context.}
\begin{equation}
\Sigma(p^2) = {1\over 96 \pi^2f^2} (p^2\Sigma_1 + 
\Sigma_2^{\rm con} + \Sigma_2^{\rm disc})\ ,
\label{eq:SELF_ENERGY}
\end{equation}
where I have explicitly
separated the contribution coming a disconnected 
propagator in the internal loop, $\Sigma_2^{\rm disc}$
(corresponding to Fig.~\ref{fig:MASS_vertex}(b)), from the other, connected
contributions in $\Sigma_2^{\rm con}$.  The  terms ``connected''
and ``disconnected'' are applicable at the QCD level: 
the disconnected diagrams contain gluon intermediate states in the meson
loop (Fig.~\ref{fig:FLOW}(d),(e),(f)).
The $U_A(1)$ symmetry guarantees
that the sum of
$\Sigma_2^{\rm con}$ and
$\Sigma_2^{\rm disc}$ is proportional to $m^2_{\pi_5}$. However,
this is not true for individual diagrams contributing to $\Sigma_2^{\rm con}$,
so I do not include the $m^2_{\pi_5}$  
factor in the definition.

For the moment, I calculate only the universal 
chiral logarithms, and ignore the analytic terms 
in the diagrams (which means the divergences are also ignored). 
The logarithms
come from the following integrals:
\begin{eqnarray}
\label{eq:INTEGRAL1}
\cI_1&\equiv &\int {d^4q\over (2\pi)^4}\; {1\over q^2 + m^2} \to 
{1\over 16 \pi^2}m^2\ln m^2 \\
\label{eq:INTEGRAL2}
\cI_2&\equiv& \int {d^4q\over (2\pi)^4}\; {q^2\over q^2 + m^2}= -m^2\cI_1
+ \int {d^4q\over (2\pi)^4} \to 
-{1\over 16 \pi^2}m^4\ln m^2 \ .
\end{eqnarray}

$\Sigma_1$ gets contributions only from 
Fig.~\ref{fig:KE_vertex}(a).  The result is
\begin{equation}
\label{eq:SIGMA1}
\Sigma_1 \to 
 -8m^2_{\pi_{\mu5}}\ln m^2_{\pi_{\mu5}} 
-8m^2_{\pi_{\mu}}\ln m^2_{\pi_{\mu}}
- \sum_B m^2_{K_B}
\ln m^2_{K_B} \ ,
\end{equation}
where no sum on  $\mu$ is implied, 
$B$ takes the 16 values 
$\{5,\mu5,\mu\nu (\mu<\nu),\mu, I\}$ as usual, 
and the arrow means that only the chiral logarithms
are included, as in \eqs{INTEGRAL1}{INTEGRAL2}.
Figure~\ref{fig:KE_vertex}(b) contributes only to
$\Sigma_2^{\rm con}$:
\begin{equation}
\label{eq:SIGMA2_KE}
\Sigma_{2,{\rm KE}}^{\rm con} \to 
 8m^4_{\pi_{\mu5}}\ln m^2_{\pi_{\mu5}}  +
8m^4_{\pi_{\mu}}\ln m^2_{\pi_{\mu}}  
+ \sum_B m^4_{K_B}
\ln m^2_{K_B} \ ,
\end{equation}
again with no sum on $\mu$ .
Note that among the pions, only
the vector and axial vector flavors contribute in 
eqs.~(\ref{eq:SIGMA1}) and (\ref{eq:SIGMA2_KE}).
It is easy to show, using the fact that $[\xi_5,I]=[\xi_5,\xi_{\mu\nu}]=
[\xi_5,\xi_5]=0$, that no terms of the form $\pi_5^4$, 
$\pi_5^2\pi^2_{\mu\nu}$, or $\pi_5^2\pi^2_I$ can be generated
by the kinetic energy term. The absence of coupling to $\pi^2_I$
explains why the kinetic energy term does not contribute
to $\Sigma_{2}^{\rm disc}$.

Figure \ref{fig:MASS_vertex}(a) 
contributes only to $\Sigma_2^{\rm con}$.  Couplings of $\pi_5$
to all the other pions are generated by the mass vertex, so there is
no restriction on the terms that can enter.  I find
\begin{eqnarray}
\label{eq:SIGMA2_MASS}
\Sigma_{2,{\rm MASS}}^{\rm con} &\to& 
-3m^4_{\pi_{5}}\ln m^2_{\pi_{5}} 
-4m^2_{\pi_5}m^2_{\pi_{\mu5}}\ln m^2_{\pi_{\mu5}}  \nonumber \\
&&-18m^2_{\pi_5}m^2_{\pi_{\mu\nu}}\ln m^2_{\pi_{\mu\nu}}  \
-4m^2_{\pi_5}m^2_{\pi_{\mu}}\ln m^2_{\pi_{\mu}}  \qquad\qquad\qquad\qquad
[{\rm no\ sum\ on\ }\mu,\nu] 
\nonumber \\
&&-3m^2_{\pi_5}m^2_{\pi_{I}}\ln m^2_{\pi_{I}} 
-(m^2_{\pi_5}+m^2_{K_5})\sum_B m^2_{K_B} \ln m^2_{K_B} \ .
\end{eqnarray}
Here, the factors of $m^2_{\pi_5}$ or $m^2_{K_5}$ in each term come from
the quark masses, which are proportional to Goldstone particle masses
squared by eq.~(\ref{eq:MASS8}) with $\Delta(\xi_5)=0$ (eq.~(\ref{eq:DELTAS})).

The mass vertex also generates the only
contributions to the disconnected diagram, 
Fig.~\ref{fig:MASS_vertex}(b).  This is the sum
of $1,2,3,\cdots$\  insertions of the vertex
${2m_0^2\over 3}
(\pi_I + S_I)^2$ on the internal $\pi_I$ line.
The $\pi_I$  line in turn comes from
a term proportional to $\pi_5^2\pi_I^2$ in the mass vertex;
the same term generates the connected contribution 
$-3m^2_{\pi_5}m^2_{\pi_{I}}\ln(m^2_{\pi_{I}})$ in
eq.~(\ref{eq:SIGMA2_MASS}). A single insertion of the $m_0^2$ vertex,
which would be the only contribution in the quenched case, gives
\begin{equation}
\label{eq:SIGMA2_DISC-QUENCH}
\Sigma_{2}^{\rm disc,quench}= 3m^2_{\pi_5}\int {d^4q \over \pi^2} {4m_0^2/3\over
(q^2 + m_{\pi_I}^2)^2} \ .
\end{equation}
The standard quenched double pole is evident.  
Iterating the $m_0^2$ vertex and summing the geometric series then
results in
\begin{equation}
\label{eq:SIGMA2_DISC}
\Sigma_{2}^{\rm disc}= 3m^2_{\pi_5}\int {d^4q \over \pi^2} {4m_0^2/3\over
(q^2 + m_{\pi_I}^2)^2\left(1 + {4m_0^2/3\over q^2 + m_{\pi_I}^2}
+ {4m_0^2/3\over q^2 + m_{S_I}^2}\right)} \ .
\end{equation}
There is no here need to take the $m_0^2\to\infty$ limit or to
perform the integral; the conversion to
the 2+1 flavor case can best be done directly from this expression.
Note that $\pi_I$ and $S_I$ intermediate states now enter.
As discussed in Sec.~\ref{sec:FLOW}, these correspond to virtual $\ell$ and $s$
quark loop contributions.

Finally, we need to consider the KS-flavor violating vertices,
eq.~(\ref{eq:V}).  It is not hard to show that,
as in the kinetic energy case, only the vector and axial pions
contribute. From Fig.~\ref{fig:V_vertex}, I obtain 
(no sum on $\mu$)
\begin{equation}
\label{eq:SIGMA2_V}
\Sigma_{2,\cV}^{\rm con} \to 
 -8a^2\Delta(\xi_{\mu5})m^2_{\pi_{\mu5}}\ln m^2_{\pi_{\mu5}}  
-8a^2\Delta(\xi_{\mu})m^2_{\pi_{\mu}}\ln m^2_{\pi_{\mu}} 
- \sum_B a^2\Delta(\xi_B) m^2_{K_B}
\ln m^2_{K_B} \ ,
\end{equation}
where $\Delta$ is given in eq.~(\ref{eq:DELTAS}).

We can now put the results together to find the $\pi_5$ mass
at one loop.  Writing 
\begin{equation}
\label{eq:MPI1loop}
(m_{\pi_5}^{\rm 1-loop})^2 = m_{\pi_5}^2 + {1\over 96\pi^2f^2}\epsilon_5\ ,
\end{equation}
we have
\begin{equation}
\label{eq:epsilon5}
\epsilon_5 = 
\Sigma_{2,{\rm KE}}^{\rm con}  +
\Sigma_{2,{\rm MASS}}^{\rm con} +
\Sigma_{2,\cV}^{\rm con} + 
\Sigma_{2}^{\rm disc}
-m_{\pi_5}^2\Sigma_1\ .
\end{equation}
From eqs.~(\ref{eq:SIGMA1}), (\ref{eq:SIGMA2_KE}), (\ref{eq:SIGMA2_MASS}), and 
(\ref{eq:SIGMA2_V})
we then have
\begin{eqnarray}
\label{eq:ans4+4}
\epsilon_5 \to  m_{\pi_5}^2 \Big(
&&-3m^2_{\pi_{5}}\ln m^2_{\pi_{5}} 
+12m^2_{\pi_{\mu5}}\ln m^2_{\pi_{\mu5}} 
-18m^2_{\pi_{\mu\nu}}\ln m^2_{\pi_{\mu\nu}}   \nonumber \\
&&\ \ +12m^2_{\pi_{\mu}}\ln m^2_{\pi_{\mu}}  
-3m^2_{\pi_{I}}\ln m^2_{\pi_{I}} \Big) + 
\Sigma_{2}^{\rm disc}
\qquad\qquad
[{\rm no\ sum\ on\ }\mu,\nu] \ .
\end{eqnarray}
where the arrow, as usual, means that only the chiral logarithms
are kept.
Note that $\epsilon_5$ is proportional to $m_{\pi_5}^2$,
as it must be
by Goldstone's theorem.
($\Sigma_{2}^{\rm disc}$ already has a factor of
$m_{\pi_5}^2$ in eq.~(\ref{eq:SIGMA2_DISC}).)  The term
$\sum_B m^4_{K_B}\ln(m^2_{K_B})$ (\eq{SIGMA2_KE})
combines with
$-\sum_B a^2\Delta(\xi_B) m^2_{K_B}\ln(m^2_{K_B})$ (\eq{SIGMA2_V})
to cancel against
$-m^2_{K_5}\sum_B m^2_{K_B} \ln(m^2_{K_B})$  (\eq{SIGMA2_MASS}).
Here I have used
\begin{equation}
m^2_{K_B} = m^2_{K_5} + a^2\Delta(\xi_B)\ ,
\label{eq:MKB}
\end{equation}
which follows from eqs.~(\ref{eq:MASS8}) and (\ref{eq:DELTAS}).
Similarly, the terms
 $8m^4_{\pi_{\mu5}}\ln(m^2_{\pi_{\mu5}})$ and 
$8m^4_{\pi_{\mu}}\ln(m^2_{\pi_{\mu}}) $ in \eq{SIGMA2_MASS} combine with corresponding
terms in \eq{SIGMA2_V} to produce terms that have a factor of $m_{\pi_5}^2$.
Indeed, \eq{SIGMA2_V}, the effect of the symmetry violating vertices, is completely
determined by the requirement that the full answer go like $m_{\pi_5}^2$ (up to logs)
in the chiral limit.  This makes it clear, for example, that the $C_i$ coefficients in
\eq{V} must enter \eq{SIGMA2_V} only through the combinations $\Delta(\xi_B)$, 
and that the final answer can depend only on the various meson masses.

Terms of the form $m_{\pi_5}^2m^2_{K_B} \ln(m^2_{K_B})$
also cancel between \eqs{SIGMA1}{SIGMA2_MASS}.
This is ``accidental'' in the sense that it is not required by symmetry.
It corresponds to  the fact that there happen to be no $\ln(m^2_{K})$ terms
in the standard continuum 2+1 flavor result \cite{GASSER_LEUTWYLER}.  For similar
quantities, such as $f_\pi$ or $\langle 0\vert  \bar u u \vert 0\rangle$, 
the $\ln(m^2_{K})$ terms will not cancel \cite{GASSER_LEUTWYLER}.

Finally, note that in the symmetry limit 
($m_{\pi_5}=m_{\pi_{\mu5}}=m_{\pi_{\mu\nu}}=m_{\pi_{\mu}}=m_{\pi_I}$)
the entire one-loop correction in \eq{ans4+4} is proportional to $\Sigma_2^{\rm disc}$.
This is also true of the standard result \cite{GASSER_LEUTWYLER}; the reason 
will be explained  in the next section.

\section{One-loop pion mass for 2+1 Dynamical Flavors}
\label{sec:FLOW}
To adjust the result in \eq{ans4+4} to the case of 2+1 dynamical flavors,
we need to identify the contribution from each of the possible quark flow
diagrams shown in Fig.~\ref{fig:FLOW}.  Many of the arguments used to
identify the quark flows will be familiar from Ref.~\cite{SHARPE_QCHPT}.
Once the contributions from virtual quark loops are determined, the adjustment
to 2+1 dynamical flavors is accomplished  by multiplying
every $\ell$ (\ie $u,d$) quark loop by $1/2$ and
every $s$ quark loop by $1/4$.

First note that Figs.~\ref{fig:FLOW}(d), (e), and (f) have disconnected internal
meson propagators, and correspond to Fig.~\ref{fig:MASS_vertex}(b). 
Figure~\ref{fig:FLOW}(d) has a single $m_0^2$ insertion and therefore
generates the contribution in \eq{SIGMA2_DISC-QUENCH}. 
Figures~\ref{fig:FLOW}(e) and (f) 
represents the iteration of the $m_0^2$ vertex through the introduction of
virtual quark loops.  Together, Figs.~\ref{fig:FLOW}(e) and (f)
change \eq{SIGMA2_DISC-QUENCH} to \eq{SIGMA2_DISC}. 

\begin{figure}[thb]
\null
\vspace{0.5truein}
\epsfxsize=0.79 \hsize
\epsffile{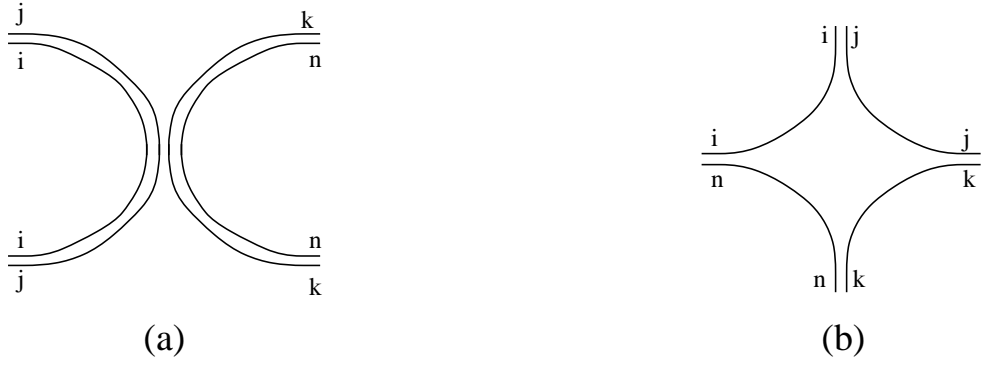}
\vspace{0.3truein}
\caption{Quark flow diagrams corresponding to 2 into 2 meson scattering
at tree level in the chiral expansion.
i,j,k,n are flavor indices.
}
\label{fig:SCAT}
\end{figure}

To identify contributions from the remaining graphs in Fig.~\ref{fig:FLOW},
we first study the possible 2 into 2 meson scattering
diagrams in the quark flow language, Fig.~\ref{fig:SCAT}. Letting $i,j,k,n=1,2,\cdots 8$
be quark flavor indices, Fig.~\ref{fig:SCAT}(a) corresponds to
a meson vertex of the form $\Phi_{ij}\Phi_{ji}\Phi_{kn}\Phi_{nk}$,
where $\Phi$ is the meson field in \eq{SIGMAandPHI88}.
Similarly Fig.~\ref{fig:SCAT}(b) corresponds to
$\Phi_{ij}\Phi_{jk}\Phi_{kn}\Phi_{ni}$.  Now,
consider the case
where all four quark flavors $i,j,k,n$ are different.  Then vertices
generated by the kinetic energy or mass terms in \eq{SU8_LAG} cannot
have the form $\Phi_{ij}\Phi_{ji}\Phi_{kn}\Phi_{nk}$ because they are
formed only by a single trace of fields (and the diagonal matrix $\cM$). 
Since Fig.~\ref{fig:SCAT}(a) cannot depend on the flavors of the quark lines
to this order
(except trivially through $\cM$),
this implies that Fig.~\ref{fig:SCAT}(a) is not generated at tree level
by the kinetic or mass terms for any values of $i,j,k,n$.  

The KS-flavor violating vertices behave differently.  Because 
at least some of the
$\xi_B$ explicitly entering \eq{V} must be off-diagonal, it is not hard
to see  that Fig.~\ref{fig:SCAT}(a) can in fact be generated.  Note, however,
that since $\xi_B$ are diagonal under the vector $SU(2)$ that mixes the
two KS fields, Fig.~\ref{fig:SCAT}(a) still vanishes when $i,j$ are of
$\ell$ type (upper $4\times4$ block) and $k,n$ are of $s$ type (lower
$4\times4$ block).

Now consider Fig.~\ref{fig:FLOW}(c).  
This graph is absent when the quarks in the virtual loops are of $s$ type
because the needed vertex vanishes to this order.
If we then choose $m_s=m_\ell$, it can make no difference
whether a quark in a virtual loop is $\ell$ or $s$ type, so this graph must also vanish
when the loop quarks are of $\ell$ type.  A slight generalization
of this argument shows that the graph again vanishes when there is one virtual quark
of each type (\ie a kaon loop).\footnote{Add a third ($s'$) KS quark field,
plus a pseudoquark KS field $\tilde s'$ of the same mass. This partially 
quenched theory is then equivalent to the original theory in the $\ell,s$
sector.  The graph Fig.~\protect{\ref{fig:FLOW}}(c) then vanishes 
when there is one $s$ and one $s'$ in the loop. Now choose
$m_{s'}=m_\ell$.}

Once we know that Fig.~\ref{fig:FLOW}(c) is absent, intermediate kaons could
only be generated by Fig.~\ref{fig:FLOW}(b).  Since the $K$ contribution to
$\epsilon_5$ in fact vanishes to this order, graphs of type
Fig.~\ref{fig:FLOW}(b) with an $s$ type quark in the virtual loop
must cancel among themselves.   Once more choosing $m_s=m_\ell$, this also implies
that Fig.~\ref{fig:FLOW}(b) cancels for a $\ell$ type virtual quark loop.

The surviving connected contribution to $\epsilon_5$ comes
from Fig.~\ref{fig:FLOW}(a). Furthermore, since the vertex in
Fig.~\ref{fig:FLOW}(a) is of the type in Fig.~\ref{fig:SCAT}(a), which
can only be generated by KS-flavor violating terms, such connected
contributions must vanish in the symmetry limit.  This is in fact the
case, as was mentioned at the end of the last section.  Because Fig.~\ref{fig:FLOW}(a)
contains valence quark lines only, these contributions do not change when
we go from 4+4 to 2+1 dynamical flavors.

The only required adjustment is therefore to the disconnected contribution,
$\Sigma_{2}^{\rm disc}$.  We need to divide every virtual quark loop 
contribution in Figs.~\ref{fig:FLOW}(e) and (f) by 2 or 4, depending on whether
it is $\ell$ or $s$ type, respectively.  In \eq{SIGMA2_DISC} this is easily accomplished
by the replacements $(4m_0^2/3)/(q^2 + m_{\pi_I}^2)\to (2m_0^2/3)/(q^2 + m_{\pi_I}^2)$
and $(4m_0^2/3)/(q^2 + m_{S_I}^2)\to 
(m_0^2/3)/(q^2 + m_{S_I}^2)$ in the denominator.
In other words, we have in the 2+1 case:
\begin{equation}
\label{eq:SIGMA2_DISC_ADJUST}
\Sigma_{2}^{\rm disc}= 3m^2_{\pi_5}\int {d^4q \over \pi^2} {4m_0^2/3\over
(q^2 + m_{\pi_I}^2)^2\left(1 + {2m_0^2/3\over q^2 + m_{\pi_I}^2}
+ {m_0^2/3\over q^2 + m_{S_I}^2}\right)} \ .
\end{equation}
We can now take the limit $m_0^2\to\infty$ and put \eq{SIGMA2_DISC_ADJUST}
in the form
\begin{equation}
\label{eq:SIGMA2_DISC_ADJUST1}
\Sigma_{2}^{\rm disc}= m^2_{\pi_5}\int {d^4q \over \pi^2} \left(
{6\over (q^2 + m_{\pi_I}^2) } - {2\over (q^2 + m_{\eta_I}^2) }
\right) \ ,
\end{equation}
where $m_{\eta_I}^2\equiv (2m_{S_I}^2 + m_{\pi_I}^2)/3$. Equation~(\ref{eq:INTEGRAL1})
then gives
\begin{equation}
\label{eq:SIGMA2_DISC_ADJUST2}
\Sigma_{2}^{\rm disc}\to m^2_{\pi_5}\left(
6 m_{\pi_I}^2\ln m_{\pi_I}^2  - 2m_{\eta_I}^2\ln m_{\eta_I}^2
\right) \ .
\end{equation}
Inserting this into \eq{ans4+4}, we get the 2+1 result for the chiral
logarithms
\begin{eqnarray}
\label{eq:ans2+1}
\epsilon_5 \to  m_{\pi_5}^2 \Big(
&&-3m^2_{\pi_{5}}\ln m^2_{\pi_{5}} 
+12m^2_{\pi_{\mu5}}\ln m^2_{\pi_{\mu5}} 
-18m^2_{\pi_{\mu\nu}}\ln m^2_{\pi_{\mu\nu}}   \nonumber \\
&&\ \ +12m^2_{\pi_{\mu}}\ln m^2_{\pi_{\mu}}  
+3m^2_{\pi_{I}}\ln m^2_{\pi_{I}}   - 2m_{\eta_I}^2\ln m_{\eta_I}^2
\Big) 
\qquad\qquad
[{\rm no\ sum\ on\ }\mu,\nu] \ .
\end{eqnarray}
In the KS-symmetry limit, this gives the standard result \cite{GASSER_LEUTWYLER}.
As discussed above, the result  
comes entirely from $\Sigma_{2}^{\rm disc}$ in this limit.

What about the analytic terms?  These can in general
come from a quite complicated set of 
operators in the chiral theory: standard $\cO(m^2)$
(``$p^4$'') operators \cite{GASSER_LEUTWYLER},
 $\cO(ma^2)$ (\eg $\cV{\rm tr}(\cM\Sigma + \cM\Sigma^\dagger)$) 
or $\cO(a^4)$ (\eg $\cV^2$) chiral operators generated by the same terms in the 
continuum effective action
that led to \eq{SU8_LAG}, and 
entirely new chiral operators of $\cO(ma^2)$ and $\cO(a^4)$
coming for example from terms in the continuum action that Lee and Sharpe 
\cite{LEE_SHARPE} call
$S_6^{\rm FF(B)}$, which break the symmetries of \eq{SU8_LAG}
down to the lattice symmetries.  However, as a function of quark mass,
the analytic terms 
in $(m^{\rm 1-loop}_{\pi_5})^2$ of the relevant order
can only be proportional to $m_\ell$, $m_\ell^2$ or $m_\ell m_s$, since they
must vanish as $m_\ell \to0$.   This means that terms of $\cO(a^4)$ in the
joint $m$, $a^2$ expansion cannot enter.

Putting \eq{ans2+1} in  \eq{MPI1loop}, adding in the analytic terms
just discussed, and writing the overall $m_{\pi_5}^2$ as $2\mu m_\ell $
(\eqs{MASS8}{DELTAS}), I arrive at the final result in the 2+1 case
(no sum on $\mu,\nu$):
\begin{eqnarray}
\label{eq:final2+1}
&&(m^{\rm 1-loop}_{\pi_5})^2/m_\ell  = 
2\mu\Bigg\{1+  {1\over 16\pi^2f^2}\Big(-\half 
m^2_{\pi_{5}}\ln {m^2_{\pi_{5}} \over \Lambda^2}
+2m^2_{\pi_{\mu5}}\ln{ m^2_{\pi_{\mu5}} \over \Lambda^2}
-3m^2_{\pi_{\mu\nu}}\ln{ m^2_{\pi_{\mu\nu}}\over \Lambda^2}   \nonumber \\
&&\ \ \ +2m^2_{\pi_{\mu}}\ln{ m^2_{\pi_{\mu}}\over \Lambda^2} 
+\half m^2_{\pi_{I}}\ln{ m^2_{\pi_{I}}\over \Lambda^2}   
-\third m_{\eta_I}^2\ln{ m_{\eta_I}^2\over \Lambda^2}
\Big) +2m_\ell K_3 + (2m_\ell  + m_s)K'_4 + a^2C\Bigg\}\ ,
\end{eqnarray}
where $\Lambda$ is the chiral scale and $K_3$, $K'_4$, and $C$ are 
independent of $m$ and $a$ to this order. 
The term $a^2C$ may alternatively be considered as a discretization correction
to $\mu$.
$K'_4$ is related to the parameter $K_4$ of Ref.~\cite{GASSER_LEUTWYLER}
by 
\begin{equation}
\label{eq:K4}
K_4 = (2m_\ell  + m_s)K'_4\ .
\end{equation}
In the KS-symmetry limit \eq{final2+1} reduces to the result in
\cite{GASSER_LEUTWYLER}.

Since the term $a^2C$ is what absorbs mass-independent cutoff dependence
in \eq{final2+1}, we
can make a rough estimate of its size by computing the change of the logarithms
when $\Lambda$ changes between two reasonable values, say 0.5 and $1.0\ \GeV$.
For MILC simulations at $a\approx 0.13$ fm, for which $\Delta_{\rm max}\approx
(400\; \MeV)^2$  (see discussion before \eq{DELTAMAX}), I find
$a^2C\sim 0.05$.
Discretization corrections to $K_3$ and $K'_4$
come from higher, $\cO(m^2a^2)$, contributions not considered
in \eq{final2+1}. 
I expect that such
corrections will be significant because they  can be generated by
the same operators in the effective QCD Lagrangian that
produced the KS-symmetry breaking potential, $\cV$.

\section{One-loop pion mass in the Quenched Case}
\label{sec:QUENCH}

Given the understanding of the quark flows in Fig.~\ref{fig:FLOW}
developed in the previous section,
the result in the quenched case is easy to write down.
$\Sigma_{2}^{\rm disc}$ in 
\eq{ans4+4} changes to the quenched version,
$\Sigma_{2}^{\rm disc,quench}$, 
which corresponds to Fig.~\ref{fig:FLOW}(d)
only. The other terms in \eq{ans4+4}, which arise from
Fig.~\ref{fig:FLOW}(a), are unaffected.
We thus have, in the quenched case
\begin{eqnarray}
\label{eq:ans_quench}
\epsilon_5 \to  m_{\pi_5}^2 \Big(
&&-3m^2_{\pi_{5}}\ln m^2_{\pi_{5}} 
+12m^2_{\pi_{\mu5}}\ln m^2_{\pi_{\mu5}} 
-18m^2_{\pi_{\mu\nu}}\ln m^2_{\pi_{\mu\nu}}   \nonumber \\
&&\ \ +12m^2_{\pi_{\mu}}\ln m^2_{\pi_{\mu}}  
-3m^2_{\pi_{I}}\ln m^2_{\pi_{I}} \Big) + 
\Sigma_{2}^{\rm disc,quench}
\qquad\qquad
[{\rm no\ sum\ on\ }\mu,\nu] \ ,
\end{eqnarray}
with $\Sigma_{2}^{\rm disc,quench}$ given by \eq{SIGMA2_DISC-QUENCH}.

Since $m_0^2$ is not to be taken to infinity in the quenched case, it is
necessary to consider other $\phi_I$ dependence in the 
Lee-Sharpe Lagrangian,\footnote{
In the quenched case,
there is no point in going to the 4+4 Lagrangian,
\eq{SU8_LAG}, since $s$ type quarks do not couple to the
pion.}
\eq{LS_LAG}. 
In Ref.~\cite{CBMG_QCHPT} it was shown that it is sufficient 
to add in a contribution to the $\phi_I$ kinetic energy:
$(2\alpha/3)(\partial_\mu\phi_I)^2$. This merely changes $m_0^2$ in
\eq{SIGMA2_DISC-QUENCH} to $m_0^2 + \alpha q^2$.

The non-analytic terms in $\Sigma_{2}^{\rm disc,quench}$ may be
extracted using
\begin{eqnarray}
\label{eq:INTEGRAL3}
\cI_3 &\equiv & \int {d^4q\over (2\pi)^4}\; {1\over (q^2 + m^2)^2}
= {-1\over 2m}{\partial\over\partial m}\cI_1 \to 
-{1\over 16 \pi^2}\ln m^2 \\
\label{eq:INTEGRAL4}
\cI_4 &\equiv & \int {d^4q\over (2\pi)^4}\; {q^2\over (q^2 + m^2)^2} 
= {-1\over 2m}{\partial\over\partial m}\cI_2 = \cI_1 - m^2\cI_3 \to 
{1\over 16 \pi^2}\;2m^2\ln m^2 \ ,
\end{eqnarray}
where $\cI_1$ and $\cI_2$ are defined in \eqs{INTEGRAL1}{INTEGRAL2}.

Putting the logarithms in $\Sigma_{2}^{\rm disc,quench}$
together   with \eq{ans_quench}, and using \eq{MPI1loop},
gives the final quenched answer:
\begin{eqnarray}
\label{eq:final_quench}
&&(m^{\rm 1-loop}_{\pi_5})^2/m_\ell  = 
2\mu\Bigg\{1+  {1\over 16\pi^2f^2}\Big(-\half 
m^2_{\pi_{5}}\ln {m^2_{\pi_{5}} \over \Lambda^2}
+2m^2_{\pi_{\mu5}}\ln{ m^2_{\pi_{\mu5}} \over \Lambda^2}
-3m^2_{\pi_{\mu\nu}}\ln{ m^2_{\pi_{\mu\nu}}\over \Lambda^2}   \nonumber \\
&&\ \ \ +2m^2_{\pi_{\mu}}\ln{ m^2_{\pi_{\mu}}\over \Lambda^2} 
-\half m^2_{\pi_{I}}\ln{ m^2_{\pi_{I}}\over \Lambda^2}   
-\twothirds m_{0}^2\ln{ m_{\pi_I}^2\over \Lambda^2}
+\fourthirds \alpha m_{\pi_I}^2\ln{ m_{\pi_I}^2\over \Lambda^2}
\Big) +2m_\ell K_3 + a^2C\Bigg\}\ ,
\end{eqnarray}
where of course the parameters $\mu$, $K_3$ and $C$ may be different from
those in \eq{final2+1}.  I keep the term multiplying $\alpha$ separate
from the other $\ln m_{\pi_I}^2$ term, since they arise from different
integrals and therefore will have different finite volume corrections.
Analytic terms depending on $m_0^2$ or $\alpha$ are not included since they 
can be absorbed in $\mu $ and $\Lambda$.
Note that the
parameter $K'_4$ does not appear here because it multiplies $m_s$,
which cannot enter in the quenched pion mass.  

In the symmetric limit, \eq{final_quench} agrees with results in
\cite{CBMG_QCHPT} or \cite{SHARPE_QCHPT}. 
As usual, the power counting in the quenched case is somewhat 
problematical; one must assume that $m_0^2$ and $\alpha$
are in some sense small to stop at one loop \cite{CBMG_QCHPT,SHARPE_QCHPT}.

\section{Finite Volume Corrections}
\label{sec:FINITE_VOLUME}

The technique for computing the finite volume corrections to
the results of Secs.~\ref{sec:FLOW} and \ref{sec:QUENCH} is standard 
\cite{FINITE_SIZE,GL_SMALL-M}. However, I will provide enough detail here to allow
the reader to include the finite volume corrections numerically in
fits to \eqs{final2+1}{final_quench}.  I assume that the Euclidean
time dimension, $T$, is large enough that it may be taken infinite,
and only corrections from the finite spatial dimensions (length $L$)
need be considered.  The generalization to the case where $T$ is also
finite will be obvious.

The results in finite volume come simply from replacing the integrals
$\cI_i$ ($i=1,\dots,4$)
in \eqsfour{INTEGRAL1}{INTEGRAL2}{INTEGRAL3}{INTEGRAL4} by their
finite volume counterparts $\cI^{(L)}_i$.
We have, for example,
\begin{eqnarray}
\label{eq:VOL-INTEGRAL1}
\cI^{(L)}_1 & \equiv& {1\over L^3}\sum_{\vec n} 
\int {dq_0\over (2\pi)}{1\over q_0^2 + (\vec q_n)^2 + m^2} \\
&& \nonumber \\
\label{eq:q_and_n}
\vec q_n &=& 2\pi\vec n/L\ ; \qquad \vec n = (n_x,n_y,n_z) 
\end{eqnarray}
where $n_x$, $n_y$, and $n_z$ are integers, and I am assuming
periodic boundary conditions.
$\cI^{(L)}_2$, $\cI^{(L)}_3$ and
$\cI^{(L)}_4$  are defined analogously.

Since the integrals of interest are divergent, it is preferable to
work with the differences $\cI^{(L)}_i-\cI_i$, which are finite
as long as care is taken in introducing a cutoff.
Examples of acceptable regulators include 
multiplying all integrands/summands by 
$(\Lambda^2/(q^2+\Lambda^2))^k$ (where $k$ is a large enough power
to render the integrals convergent) \cite{GOLTERMAN_LEUNG}, 
point-splitting the corresponding position-space propagators,
and defining the theory on a lattice.

Assuming the integrals/sums have been regulated appropriately,
I define the dimensionless finite volume corrections $\delta_i$ by
\begin{eqnarray}
\label{eq:delta1_defn}
\cI^{(L)}_1 - \cI_1 &=& {m^2\over 16 \pi^2}\; \delta_1(mL) \\
\label{eq:delta2_defn}
\cI^{(L)}_2 - \cI_2 &=& {m^4\over 16 \pi^2}\; \delta_2(mL) \\
\label{eq:delta3_defn}
\cI^{(L)}_3 - \cI_3 &=& {1\over 16 \pi^2}\; \delta_3(mL) \\
\label{eq:delta4_defn}
\cI^{(L)}_4 - \cI_4 &=& {m^2\over 16 \pi^2}\; \delta_4(mL) 
\end{eqnarray}

The correction $\delta_1$ 
can easily be written in terms of the Euclidean position space 
propagator, defined in infinite volume by
\begin{equation}
\label{eq:G} 
G(x) \equiv 
\int {d^4q\over (2\pi)^4}\; {e^{-iq\cdot x}\over q^2 + m^2} = 
{m\over 4\pi^2 |x|} K_1(m|x|)\ ,
\end{equation}
where $K_1$ is a Bessel function of imaginary argument.
The corresponding finite volume propagator, is given by
\begin{equation}
\label{eq:GL}
G^{(L)}(x) \equiv 
{1\over L^3}\sum_{\vec n}
\int {dq_0\over (2\pi)}{e^{-iq\cdot x}\over q_0^2 + (\vec q_n)^2+ m^2} = 
G(x) + \sum_{\vec n\not= 0} G(x + L\vec n)
\end{equation}
with $q = (q_0,\vec q_n)$.  The last equality follows from the Poisson
resummation formula, or simply by noting that it is the unique solution
of the corresponding differential equation with the correct (periodic)
boundary conditions.
Subtracting \eqs{GL}{G}, setting $x=0$, and putting
the result into \eq{delta1_defn} using \eqs{VOL-INTEGRAL1}{INTEGRAL1}
gives
\begin{equation}
\label{eq:delta1}
\delta_1(mL) =  {4\over mL}\sum_{\vec n\not= 0}{K_1(|\vec n|mL)\over |\vec n|}
\ .
\end{equation} 
For practical values of $mL$, $\delta_1$ can be readily computed to machine
precision with this formula. (One is limited of course by the accuracy
to which $K_1$ is evaluated.)  An alternative approach, inspired by
Ref.~\cite{NOBES}, is to compute the difference of the (regulated) sum
and integral directly in momentum space, treating the sum as
an integral by rounding $L\vec q/(2\pi)$ to the nearest $\vec n$ and using a 
standard numerical integration program \cite{Lepage:1980dq}.
With a reasonable regulator (\eg the one from \cite{GOLTERMAN_LEUNG},
mentioned above), this method gives results consistent with
\eq{delta1}, but it is much slower.

The other finite volume corrections follow from the relations among the
integrals given in \eqsthree{INTEGRAL2}{INTEGRAL3}{INTEGRAL4} and
the standard recursion relation $K'_1(z) = -K_1(z)/z - K_0(z)$,
where prime implies differentiation. I find
\begin{eqnarray}
\label{eq:delta2}
\delta_2(mL) & = &  -\delta_1(mL) \\
\label{eq:delta3}
\delta_3(mL) & = &  -\delta_1(mL)-{mL\over2}\delta'_1(mL) = 2 \sum_{\vec n\not= 0}K_0(|\vec n|mL) \\
\label{eq:delta4}
\delta_4(mL) & = &  2\delta_1(mL)+{mL\over2}\delta'_1(mL) = 
\delta_1(mL)-\delta_3(mL) 
\end{eqnarray} 
In deriving \eq{delta2}, one must be careful to regulate so that the difference
between the sum and the integral of unity vanishes in the limit of infinite
regulator. The regulator in \cite{GOLTERMAN_LEUNG} works well, for example.

Large $mL$ asymptotic expressions for the $\delta_i$ are obtained by including
only the 6 nearest neighbors in the sums in \eqs{delta1}{delta3}
and using the only leading terms in $K_1$ and $K_0$. I find
\begin{eqnarray}
\label{eq:delta12_asymptotic}
\delta_1(mL) & = &  -\delta_2(mL) \sim {12\sqrt{2\pi}\over (mL)^{3\over2}}
e^{-mL} \\
\label{eq:delta34_asymptotic}
\delta_3(mL) & \sim &  -\delta_4(mL) \sim 6\sqrt{2\pi\over mL}
e^{-mL}\ .
\end{eqnarray} 
While these expressions are useful for estimating the size of finite size
effects, 
the leading corrections 
to \eqs{delta12_asymptotic}{delta34_asymptotic}
have just one higher power of $1/(mL)$.
In the quantitative applications I have in mind \cite{MILC_FITS},
for which $mL$ is in the range 3 to 9,
the full expressions 
(\eqsfour{delta1}{delta2}{delta3}{delta4}) --- or at least the full
$K_1$ and $K_0$ from the nearest neighbors --- will be needed.
It is important to note, however, that for very small quark
masses, $mL\ltwid 1$, \eqsfour{delta1}{delta2}{delta3}{delta4}
are not applicable, no matter how many terms are included
in the sums.  In that limit, the zero-mode (the spatially independent
component of the pion field) must be treated exactly \cite{GL_SMALL-M}.

Comparing eqs.~(\ref{eq:delta1_defn})--(\ref{eq:delta4_defn})
with \eqsfour{INTEGRAL1}{INTEGRAL2}{INTEGRAL3}{INTEGRAL4} then gives the
rules for including finite volume corrections. In the 2+1 case,
one needs merely to make the replacement
\begin{equation}
\label{eq:finiteV2+1}
\ln {m^2\over\Lambda^2} \to \ln {m^2\over\Lambda^2} + \delta_1(mL)
\end{equation}
for every logarithm in \eq{final2+1}.
It is interesting to note that 
when $m_\ell\to0$ (with $L$ and the 
symmetry violating parameters, $C_i$, fixed),  the leading finite size
correction to \eq{final2+1} is negative, due to the
$-\half m^2_{\pi_5}\ln m^2_{\pi_5}$ term.  In the standard case,
the leading correction is positive, due to the positive sign of
$m^2_\pi\ln m^2_\pi$.

In the quenched approximation, the replacements in \eq{final_quench} are:
\begin{eqnarray}
m_{0}^2\ln{ m_{\pi_I}^2\over \Lambda^2} &\to &
m_{0}^2\ln{ m_{\pi_I}^2\over \Lambda^2} - m_0^2\delta_3(m_{\pi_I}L) \nonumber \\
\label{eq:finiteV_quench}
\alpha \ln{ m_{\pi_I}^2\over \Lambda^2} &\to &
\alpha \ln{ m_{\pi_I}^2\over \Lambda^2}+ \half\alpha\delta_4(m_{\pi_I}L)\\
\ln {m^2\over\Lambda^2} &\to& \ln {m^2\over\Lambda^2} + \delta_1(mL) 
\qquad {\rm [all\ other\ logarithms]}\ .\nonumber
\end{eqnarray}

\section{Remarks and Conclusions}
\label{sec:CONCLUSION}
My main results are given for the 2+1 flavor case in \eq{final2+1}
and for the quenched case in \eq{final_quench}. Finite volume
corrections to these formulas appear in \eqs{finiteV2+1}{finiteV_quench},
respectively, with the values of $\delta_i$ given in eqs.~(\ref{eq:delta1})
to (\ref{eq:delta4}).

\begin{figure}[thb]
\null
\vspace{0.5truein}
\epsfxsize=0.8 \hsize
\epsffile{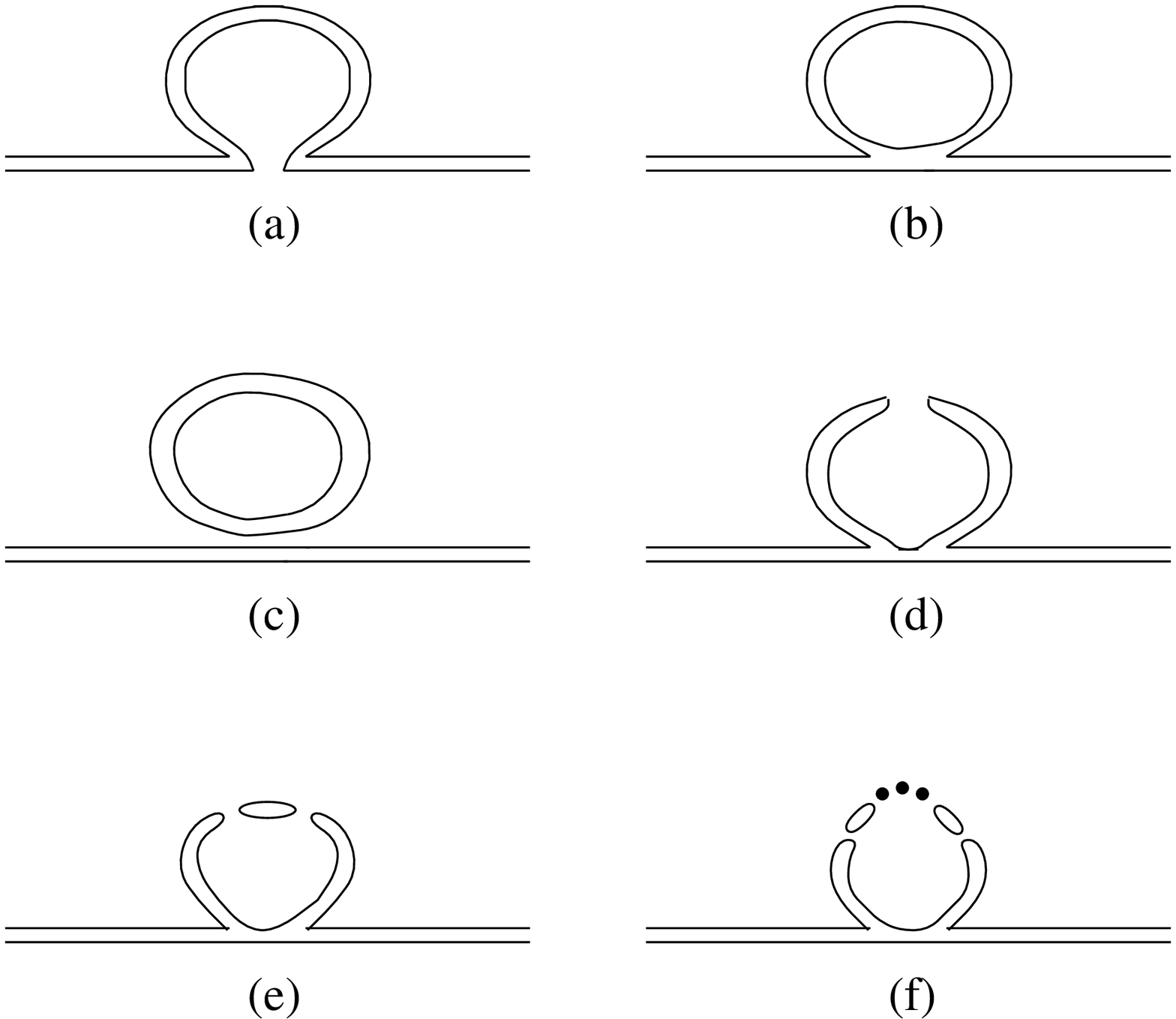}
\vspace{0.3truein}
\caption{}{Quark flow diagrams that could contribute to
the (flavor non-singlet) pion propagator at one loop in the chiral expansion. 
Diagrams (d), (e) and (f)
are disconnected graphs that correspond to figure 
Fig.~\ref{fig:MASS_vertex}(b);
(f) represents the sum over two or more intermediate virtual quark loops.
Note that (d), (e) and (f) each have a second version, in which the
two sides of the meson loop come from {\it different} valence lines.  For the
purposes of this paper,  one may consider the diagrams shown to be generic,
subsuming both versions.  Diagrams with ``$\Phi_0$ vertices,'' such
as Fig.~3(i) of Ref.~\cite{SHARPE_QCHPT}, are not relevant: either 
because they are eliminated by the $m_0^2\to\infty$ limit (the 2+1 case),
or because they have virtual quark loops (the quenched case).
Diagrams with odd numbers of $\Phi_0$ lines at a vertex, such 
as Fig.~3(f) of Ref.~\cite{SHARPE_QCHPT}, can be
eliminated by a field redefinition \cite{CBMG_QCHPT}.
}
\label{fig:FLOW}
\end{figure}

I have computed only $m^2_{\pi_5}$, but a generalization of this
calculation to $m^2_{K_5}$ is straightforward and is
in progress \cite{AUBIN_BERNARD}.
Slightly  more complicated, but also on the list of things to
compute \cite{AUBIN_BERNARD} are 
$f_\pi$, $f_K$, $\langle \bar \Psi \Psi\rangle$, and
the non-Goldstone $\pi$ and $K$ masses. 
These quantities do not vanish in the chiral limit, so there will
be fewer short-cuts available.  A further complication for
a quantity like $f_\pi$ is that there is no accidental cancellation of
diagrams corresponding to Fig.~\ref{fig:FLOW}(b).  An additional adjustment
for the number of flavors in virtual quark  loops will therefore
be required.  However, the adjustment is not difficult, since 
$\ln m_K^2$ contributions come uniquely from this diagram.
To generate the 2+1 case, one will merely have to: (1) multiply the 
$\ln m_K^2$ terms by $1/4$ to count the $s$ quarks in the loop correctly,
and (2) subtract half the (original)
$\ln m_K^2$ terms, after putting $m_s\to m_\ell$,
to count $\ell$ quarks in the loop correctly.  For quantities such as 
$f_K$ or $m_K^2$, where there will be valence $\ln m_K^2$ contributions,
the $\ln m_\pi^2$ and $\ln m_S^2$ terms (from the {\it connected}
internal propagator) can be used to make the adjustment.

Preliminary fits of \eq{final2+1} to MILC 2+1 data
are encouraging: the fits give good confidence levels
for reasonable ranges of quark mass \cite{MILC_FITS}.  The contrast with the 
standard fits that do not take into account KS flavor breaking
\cite{MILC_SPECTRUM} is especially striking. 
I emphasize here that the new fits have the same number of free parameters 
as the standard ones.  This is because the various pion splittings 
are not left free in \eq{final2+1} but are determined first by the
data for the masses of all the pions.

In the quenched case,
one generally treats $m_0$ (or equivalently
$\delta\equiv m_0^2/(24\pi^2 f^2)$) as a free parameter.  As such,
decent fits can apparently
be performed whether or not one takes flavor breaking into
account.  However, the values of $\delta$ obtained seem to be
considerably closer to the real world values ($\delta\approx0.18$)
when flavor breaking effects are included. In both the 2+1 and quenched
cases, we are collecting additional lattice data at small quark
mass \cite{MILC_FITS}, which should significantly increase the
discrimination between the symmetric and flavor-breaking fits.

\bigskip
\bigskip
\centerline{\bf ACKNOWLEDGMENTS}
\bigskip
I am grateful to M.\ Golterman, P.\ Hasenfratz, 
K.\ Orginos, S.\ Sharpe, and D.\ Toussaint,
as well as my other colleagues in MILC, for very useful
discussions. M.\ Golterman, S.\ Sharpe, and D.\ Toussaint
also provided helpful comments on an earlier draft of this manuscript. 
Some of this work was completed during the program
``Lattice QCD and Hadron Phenomenology'' 
at the Institute for Nuclear Theory, 
University of Washington (Fall, 2001).  I thank
the INT staff and
organizers M.\ Golterman and S.\ Sharpe for
a productive stay.

\vfill\eject
\null
\vspace{0.5truein}
\vskip 1truein

\vfill\eject

\vfill\eject

\vfill\eject

\begin{references}
\bibitem{ORGINOS} 
It was K.\ Orginos who first suggested to me that KS-flavor violation
might be directly responsible for
the difficulties with chiral fits in Ref.~\cite{MILC_SPECTRUM} and that a
one-loop calculation using the Lagrangian of
Ref.~\cite{LEE_SHARPE} might be necessary.
\bibitem{HASENFRATZ} P.\ Hasenfratz points out (private
communication) that one should always be
able to write down a standard (\`a la Ref.~\cite{GASSER_LEUTWYLER})
2+1 flavor chiral theory, and add arbitrary symmetry violating terms
to represent $\cO(a^2)$ effects.  But without a corresponding quark
level Lagrangian, I see no way to organize and limit these terms
to get out useful information.
\end{references}
\end{document}